\documentclass[runningheads]{svmult}
\usepackage{makeidx}   
\usepackage{graphicx}  
\usepackage{subeqnar}  
\usepackage{multicol}  
\usepackage{cropmark} 
\usepackage{physprbb}  
\begin{document}
\title*{Dim Matter in the Disks of \protect\newline Low    
Surface Brightness Galaxies}
\toctitle{Dim Matter in the Disks of
\protect\newline Low Surface Brightness Galaxies}
%
%
\titlerunning{Dim Matter in the Disks of LSBGs}
%
\author{Burkhard Fuchs}
\authorrunning{Burkhard Fuchs}
%
%
\institute{Astronomisches Rechen--Institut, 69120 Heidelberg, Germany}

\maketitle              

\begin{abstract}
An attempt is made to set constraints on the otherwise ambiguous decomposition
of the rotation curves of low surface brightness galaxies into contributions due
to the various components of the galaxies. For this purpose galaxies are 
selected which show clear spiral structure. Arguments of density wave theory of
galactic spiral arms are then used to estimate the masses of the galactic disks.
These estimates seem to indicate that the disks of low surface brightness 
galaxies might be much more massive than currently thought. This unexpected 
result contradicts stellar population synthesis models. This would also mean 
that low surface brightness galaxies are not dominated by dark matter in their 
inner parts.
\end{abstract}

\section{Introduction}
The rotation curves of low surface brightness galaxies (LSBGs) seem to indicate
that these galaxies are dominated by dark matter even in their inner parts. -- 
An LSBG is usually defined as a spiral galaxy with a central surface brightness
$\geq$ 1 mag/arcsec$^2$ fainter than a `normal' high surface brightness galaxy.
-- However, the disentanglement of the disk and dark halo contributions to the
observed rotation curves is notoriously ambiguous. Further dynamical
constraints on the decompositions of the rotation curves are therefore required.
If, for example, a galaxy shows clear spiral structure, arguments from density
wave theory of galactic spiral arms may provide such constraints (Athanassoula,
Bosma \& Papaioaunou 1987). In the following I present an analysis of a small
sample of such LSBGs.

\section{Rotation Curves of LSBGs}
Recently McGaugh, Rubin \& de Blok (2001) have measured high--resolution
rotation curves of a large set of LSBGs which are well suited to study details
of the rotation curves. If surface photometry of the galaxies was available in
the literature, the same authors (de Blok, McGaugh \& Rubin 2001) have
constructed dynamical models of the galaxies. The observed rotation curves are
modeled as
\begin{equation}
  v_{\rm c}^2(R)= v_{\rm c,bulge}^2(R)+v_{\rm c,disk}^2(R)+
  v_{\rm c,isgas}^2(R)+v_{\rm c,halo}^2(R)\,,
\end{equation}
where $v_{\rm c,bulge}$, $v_{\rm c,disk}$, $v_{\rm c,isgas}$, and 
$v_{\rm c,halo}$ denote the contributions due to the bulge, the stellar disk, 
the interstellar gas, and the dark halo, respectively. The radial variations of
$v_{\rm c,bulge}(R)$, $v_{\rm c,disk}(R)$, and $v_{\rm c,isgas}(R)$ were derived
from the observations, while the normalizations by the mass--to--light ratios
were left as free parameters of the fits of the mass models to the data. For
the dark halo de Blok, McGaugh \& Rubin (2001) argue in favour of the model of 
a quasi--isothermal sphere. They
demonstrate once again the degeneracy of the problem by providing for each
galaxy several models, one with zero bulge and disk mass, one model with a
`reasonable' mass--to--light ratio of the bulge and the disk, and finally a
`maximum--disk' model with bulge and disk masses at the maximum allowed by the
data. All fits to the observed rotation curves are of the same quality, but
de Blok, McGaugh \& Rubin (2001) noticed that in the maximum disk models the
mass--to--light ratios of the disks are unrealistically high in view of stellar
population synthesis modeling (Bell \& de Jong 2001). As an example the maximum
disk model of the galaxy UGC\,6614 is reproduced in Fig.~1. Fig.~2 shows an 
image of the galaxy reproduced from de Blok, van der Hulst \& Bothun (1995).

\begin{figure}[htb]
\begin{center}
\includegraphics[width=.7\textwidth]{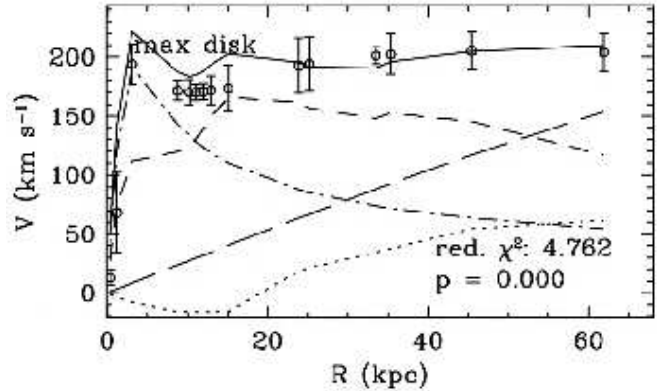}
\end{center}
\caption[]{`Maximum disk' model (solid line) of the rotation curve of UGC\,6614 
according to de Blok, McGaugh \& Rubin (2001). The dash--dotted line indicates
the bulge contribution, the dashed line the disk contribution, the dotted line 
the contribution by the interstellar gas, and the long--dashed line the dark 
halo contribution, respectively. The radial distance scale is based on a Hubble
constant of H$_0$=75 km/s/Mpc.}
\label{fig1}
\end{figure}

\begin{figure}[htb]
\begin{center}
\includegraphics[width=.4\textwidth]{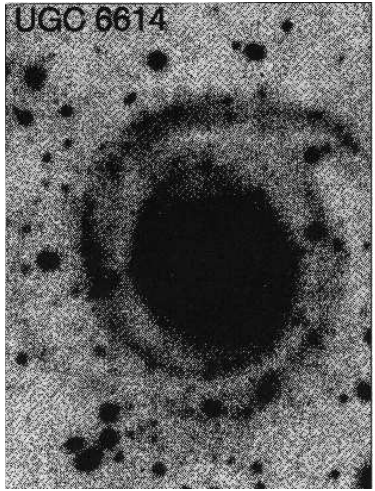}
\end{center}
\caption[]{Image of UGC\,6614 (3.4' $\times$ 3.4').}
\label{fig2}
\end{figure}

\section{Dynamical Constraints}
Since I attempt to set constraints on the decompositions of the rotation curves
of LSBGs using arguments from the density wave theory of galactic spiral arms, I
have inspected images of each galaxy from the list of McGaugh, Rubin \& 
de Blok (2001) for discernible spiral structure. The images are either
reproduced in de Blok, van der Hulst \& Bothun (1995) or were retrieved from the
Digitized Sky Survey using the Aladin data base. Further rotation curves of
LSBGs are presented in van der Hulst et al.~(1993) and Pickering et al.~(1997).
These have been obtained by HI observations and have due to beam smearing a
lower resolution than the optical rotation curves of McGaugh, Rubin \& 
de Blok (2001). The bulge regions of the galaxies are not resolved, but 
the rotation curves in the disk regions are resolved sufficiently
well and I have examined also the images of these galaxies. The resulting sample
of LSBGs, which can be used for the present purposes, is listed in Table 1
where I have also included UGC\,128 (de Blok \& McGaugh 1996). The
second column gives the observed peaks of the rotation curves, while the third
column gives the peaks of the disk contributions to the rotation curves
according to maximum disk models. The corresponding mass--to--light ratios in
the R--band are given in the fourth column. All selected galaxies show 
two--armed spiral structure ($m$ = 2).

\begin{table}[htb]
\caption{Sample of LSBGs}
\begin{center}
\begin{tabular}{lccc}\hline
   & $v_{\rm max}$ & $v_{\rm disk,max}$ & M/L$_{\rm R}$\\ \hline \hline
F\,568-1    &  140 & 94 & 9 \\
F\,568-3    &  100 & 50 & 2 \\
F\,568-6    &  300 & 86$^*$ & 1.2 \\
F\,568-V1   &  113 & 83 & 14 \\
UGC\,128    &  131 & 62 & 3 \\
UGC\,1230   &  103 & 75 & 6 \\
UGC\,6614   &  210 & 160 & 8 \\ \hline
            & km/s & km/s & M/L$_{R\odot}$\\ \hline
\end{tabular}
\vskip 0.15mm
$^*$ not a maximum disk model
\end{center}
\end{table}

The dynamics of galactic spiral structure is theoretically well understood in
the framework of density wave theory of spiral arms. Density wave theory makes,
in particular, a specific prediction for the number of spiral arms. Spiral
density waves develop in galactic disks preferentially with a circumferential
wavelength of about (Toomre 1981, Fuchs 2001, 2002)
\begin{equation}
\lambda \approx X(\frac{A}{\Omega_0}) \lambda_{\rm crit} \,,
\end{equation}
where $\lambda_{\rm crit}$ denotes the critical wavelength
\begin{equation}
\lambda_{\rm crit} = \frac{4 \pi^2 G \Sigma_{\rm d}}{\kappa^2}\,.
\end{equation}
$\Sigma_{\rm d}$ is the surface density of the disk and $\kappa$ is the
epicyclic frequency of the stellar orbits,
\begin{equation}
\kappa = \sqrt{2}\,\frac{v_{\rm c}}{R}\,\sqrt{1+\frac{R}{v_{\rm c}}
\frac{d v_{\rm c}}{d R}} \,.
\end{equation}
$G$ denotes the constant of gravitation. The coefficient $X(\frac{A}{\Omega_0})$
depends on the slope of the rotation curve measured by Oort's constant $A$,
\begin{equation}
\frac{A}{\Omega_0} = \frac{1}{2} \left( 1 - \frac{R}{v_{\rm c}}
\frac{d v_{\rm c}}{d R} \right) \,,
\end{equation}
and has been determined explicitely for various cases by Toomre (1981),
Athanassoula (1984), or Fuchs (2001). For a flat rotation curve the value is
$X(0.5) = 2$. 

\begin{figure}[htb]
\begin{center}
\includegraphics[width=.4\textwidth]{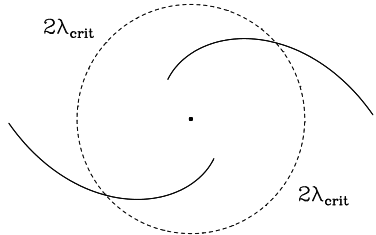}
\end{center}
\caption[]{Sketch of the spacing of spiral arms.}
\label{fig3}
\end{figure}

\noindent
As sketched in Fig.~3 the number of spiral arms is obviously determined by how
often the wavelength $\lambda$ fits onto the annulus,
\begin{equation}
m \approx \frac{2 \pi R}{X \lambda_{\rm crit}}\,.
\end{equation}
The predicted number of spiral arms (6) is based on the local model of a
shearing sheet which describes the dynamics of a patch of a galactic disk
(Goldreich \& Lynden--Bell 1965, Julian \&Toomre 1966, Fuchs 2001). Equation (6)
can be applied globally to an entire disk strictly only in the case of a
Mestel disk, which has an exactly flat rotation curve, $v_{\rm c}(R) = const.$,
and a surface density distribution of the form $\Sigma_{\rm d} = v_{\rm
c,disk}^2 /2 \pi G R$. A two--armed spiral corresponds to $v_{\rm c,disk} =
v_{\rm c}/\sqrt{2}$. In the more complicated disk models used here $m$ varies
formally with galactocentric distance as illustrated for UGC\,6614 in Fig.~4.
However, the maximum growth factor of the amplitudes of the density
waves is not sharply peaked at the circumferential wavelength (2) (Toomre 1981,
Fuchs 2001) so that density waves with smaller or greater wavelengths can
develop as well. Allowing for these side fringes of wavelengths and the
corresponding variations of the coefficient $X$ in equation (6) one can derive
for the distance range in the galactic disk spanned by the spiral arms a 
uniform value of $m$.

\begin{figure}[htb]
\begin{center}
\includegraphics[width=.5\textwidth]{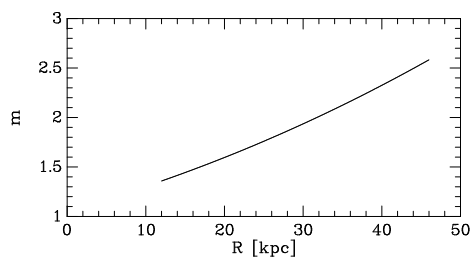}
\end{center}
\caption[]{Radial variation of the expected number of spiral arms 
in the disk of UGC\,6614.}
\label{fig4}
\end{figure}

Thus, if the rotation curve is known, one can deduce from the number of spiral
arms by inverting equation (6) the critical wavelength and from that the surface
density in absolute units. Following this prescription I have determined for
each galaxy in the sample from its rotation curve the shear rate (5)
representative for the region spanned by the spiral arms, and have derived the
dynamical mass estimates of the disks of the LSBGs summarized in Table 2. As
can be seen from the resulting mass--to--light ratios in Table 2 the
dynamical mass estimates {\it confirm} essentially the maximum disk models 
given in Table 1. 

\begin{table}[htb]
\caption{}
\begin{center}
\begin{tabular}{lccccc}\hline
   & $v_{\rm max}$ & $v_{\rm disk,max}$ & $M_{\rm disk}$ & $ M_{\rm is gas}$ &
    M/L$_{\rm R}$\\ \hline \hline
F\,568-1    &  140 & 115 & 42 & 3 & 14 \\
F\,568-3    &  100 & 87 & 18 & 2 & 7 \\
F\,568-6    &  300 & 260 & 740 & 50 & 11 \\
F\,568-V1   &  113 & 89 & 15 & 2 & 16 \\
UGC\,128    &  131 & 102 & 43 & 9 & 4 \\
UGC\,1230   &  103 & 77 & 16 & 8 & 6 \\
UGC\,6614   &  210 & 160 & 130 & 35 & 8 \\ \hline
            & km/s & km/s & $10^9 M_\odot$ & $10^9 M_\odot$ &
	     M/L$_{R\odot}$\\ \hline
\end{tabular}
\end{center}
\end{table}

\section{Discussion and Conclusions}
The high mass--to--light ratios derived dynamically contradict stellar 
population synthesis models of LSBGs. These models are based on the 
initial mass function and an adopted star formation history, conventionally
parameterized by an exponential function. The temporal evolution of the models
is usually traced by their colour. Bell \& de Jong (2001) have compared recently
models by various authors and find that they agree very consistently. The LSBGs
in the sample analyzed here have colours B -- R $\leq$ 1. The stellar population
synthesis models indicate then that galactic disks with such colours and with
ages of about 10 Gyrs should have mass--to--light ratios of $M/L_{\rm R} \leq 1 
M_\odot /L_{{\rm R}\odot}$. The star formation rate in the models is nearly
constant, and the general opinion is that the disks of LSBGs are slowly burning,
hardly evolved disks (van den Hoek et al.~2000).  

As a test that the density wave theory arguments to constrain the disk masses
are not grossly misleading I have analyzed the numerical simulation by Sellwood
(1981) of the evolution of a self gravitating disk in exactly the same way as
the LSBGs. From snapshots of the simulation and the adopted rotation curve I can
reproduce consistently the disk mass used for the simulation. I conclude
from this tentatively that the disks of LSBGs might be indeed more massive than
previously thought and contain a hitherto unknown dim component.

\begin{figure}[htb]
\begin{center}
\includegraphics[width=.7\textwidth]{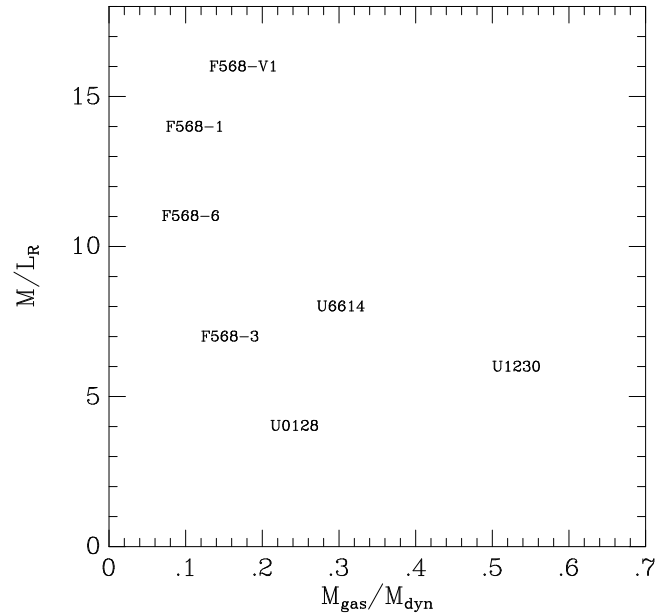}
\end{center}
\caption[]{Mass--to--light ratios of the LSBGs versus their relative gas mass
fraction.}
\label{fig5}
\end{figure}

The nature of the dim disk
component, if its existence can be confirmed, is unclear. The mass of the
interstellar gas disk is included in the dynamical mass estimate, because  
galactic spiral arms contain a lot of interstellar gas. In the fifth
column of Table 2 the gas masses of the sample galaxies compiled from the
literature (van der Hulst et al.~1993, McGaugh \& van der Hulst 1996, Pickering
et al.~1997) are listed. As can be seen from Table 2 also the inclusion of 
the interstellar gas cannot explain the high dynamical disk mass estimates. I
find a loose correlation of the mass--to--light ratios with the relative gas
mass fractions $M_{\rm isgas}/M_{\rm dyn}$ as shown in Fig.~5. This might
indicate that the galaxies with the higher mass--to--light ratios have consumed
more of their interstellar gas by star formation. The unknown disk component
might be then a dim, probably old stellar disk population. The high dynamical
disk mass estimates imply, on the other hand, that LSBGs might be less dark
matter dominated than currently thought.

There are a number of galaxies in the list of McGaugh, Rubin \& de Blok (2001)
for which no surface photometry is presently available, but which show clear
spiral structures. Work is in progress to provide uncalibrated surface
photometry for these galaxies in order to increase the sample of LSBGs for which
disk masses can be estimated by the method described here.

\subsection*{Acknowledgments}
This research has been carried out using the Simbad data base at CDS.

%

\end{document}